\documentclass{menu97}    
\usepackage{epsfig}

\begin{document}
    \setlength{\baselineskip}{2.6ex}

\title{The $\pi NN$ coupling constant from np
charge exchange scattering}

\author{B. LOISEAU\\
{\em  Division de Physique Th\'eorique\thanks{Unit\'{e} de 
Recherche des Universit\'{e}s Paris 11 et Paris 6 Associ\'{e}e au CNRS},
Institut de Physique Nucl\'eaire, 91406 Orsay Cedex and\\ LPTPE
Universit\'e P. \& M. Curie, 4 Place Jussieu, 75252 Paris Cedex 05, 
France}\\
\vspace{0.3cm}
T.E.O. ERICSON\\
{\em European Organization for 
Nuclear Research, CH-1211 Geneva 23,
Switzerland,\\
and The Svedberg Laboratory, Uppsala University, Box 533, 
S-75121 Uppsala, Sweden}\\
\vspace{0.3cm}
and\\
\vspace{0.3cm}
 J. RAHM, J. BLOMGREN, N. OLSSON\\
{\em Department of Neutron Research, Uppsala University, Box 535, 
S-75121 Uppsala, Sweden}}

\maketitle

\begin{abstract}
\setlength{\baselineskip}{2.6ex}
A novel extrapolation method has been used to deduce the charged
$\pi N\! N$ coupling constant from backward  $np$ differential  scattering
cross sections. We applied it to new measurements performed at
162~MeV  at the The Svedberg Laboratory in Uppsala.
In the angular range $150^\circ-180^\circ$, the carefully normalized 
data are steeper than those of most previous measurements. 
The extracted value, $g^2_{\pi^\pm} = 14.52 \pm 0.26$, in good agreement
with the classical value, is higher than those 
determined in recent nucleon-nucleon partial-wave analyses.
\end{abstract}

\setlength{\baselineskip}{2.6ex}

\section*{INTRODUCTION}

The knowledge of the precise value of the $\pi NN$ coupling is a crucial
issue: not only in nuclear physics where it is a fundamental constant, but
also in particle physics where it is of great importance for the
understanding of chiral symmetry breaking~\cite{Eri93}.  Its experimental
error is the main obstacle in the accurate testing of the corrections to the
Goldberger-Treiman relation as predicted from chiral symmetry breaking. With
the latest value for the axial coupling constant,
$g_A = 1.266\pm0.004$~\cite{Gre95},
this relation would lead to $g^2(q^2=0)=13.16\pm0.16$,
if it were exact, which is not expected. The uncertainty, here of
about $ \pm 1\%$, comes from the experimental error in $g_A$ and $f_{\pi}$.
If we know how to calculate the corrections perfectly we can clearly make
good use of a precision of 1\% in $g^2$.

In the 1980's, the $\pi N\! N$ coupling constant was believed to be well
known. Koch and Pietarinen~\cite{Koc80} determined a value of the charged
pion coupling constant, $g^{2}_{\pi^{\pm}} = 14.28 \pm 0.18$, from
$\pi^{\pm} p$ scattering data. Kroll~\cite{Kro81} found the neutral pion
coupling constant to be $g^{2}_{\pi^{0}} = 14.52 \pm 0.40$, analysing $pp$
data with forward dispersion relations. In the early 1990's the Nijmegen
group~\cite{Ber90,Klo91,Sto93a} determined smaller values on the basis of
energy-dependent partial-wave analyses (PWA) of nucleon-nucleon ($N\! N$)
scattering data. They obtained $g^{2}_{\pi^{0}} = 13.47 \pm 0.11$ and\linebreak
$g^{2}_{\pi^{\pm}} = 13.58 \pm 0.05$. Similar values around $g^2_\pi =13.7$,
have also been found by the Virginia Tech
group~\cite{Arn90,Arn94a,Arn94b} from analysis of both $\pi^{\pm} N$ and
$N\! N$ data. These results have stimulated an intense debate, 
and it has  become urgent to determine $g^2$ to high precision,
convincingly and model-independently~\cite{Eri93}.

In the analysis by the Nijmegen group~\cite{Sto93a} the determination of the
coupling constant seems not to be very sensitive to the backward $np$ cross
section. In our work at 162~MeV~\cite{Eri95}, we have shown the contrary
using `pseudodata' built from models in common use, including the Nijmegen
potential~\cite{Sto94}. The experimental normalization of the cross section
is crucial and this has been a well known problem in the past. Most
energy-dependent PWA's have therefore chosen to let the normalization of
data float more or less freely. The direct sensitivity to the $np$ cross
section is then lost, and the coupling constant can depend diffusely on many
observables. We believe that precision data of the backward $np$ cross section
should be one of the best places in the $N\! N$ sector to determine the
charged coupling constant.  We have demonstrated recently~\cite{Eri95,Eri96a},
that it is both the {\em shape} of the angular distribution at the most
backward angles, and the {\em absolute normalization} of the data, that are
of decisive importance in this context. Before analysing the new 162 MeV
data we shall illustrate the possible consequences of the relativly large
spread of the values of the $\pi NN$ coupling constant on the quark mass
ratio $m_s/\hat m=2 m_s/(m_u+m_d)$. We shall then describe the extrapolation
to the pion pole and give some conclusions.

\section*{DASHEN-WEINSTEIN SUM RULE}

As shown in Ref.~\cite{Fuchs90} the Dashen-Weinstein sum
rule~\cite{Dashen69} for deviations from the Goldberger-Treiman (GT)
relation allows a possible determination of $m_s/\hat m$. This sum rule is
obtained by considering the matrix elements of the non-strange charged axial
vector current at $q^2=0$ between $n$ and $p$ and the strange ones between
$p$ and $\Lambda$ and between  $n$ and $\Sigma$. The hypothesis of partial
conservation of the current leads at $q^2=0$ to equations between the
deviations of the GT relations, the $\pi NN$, $K \Lambda N$, $K \Sigma N$
couplings, the axial form factors and the pseudoscalar meson decay constants
$f_\pi$ and $f_K$. Expanding matrix elements of the pseudoscalar densities
in terms of the quark masses and using the $SU(3)_V$ symmetry invariance in
the Chiral limit, lead then to the sum rule. It expresses the ratio
$m_s/\hat m$ in terms of experimental quantities. So far one knows only
an upper limit of 7 for the $K \Sigma N$ coupling. From the present 
knowledge for the other constants, the sum rule leads then to a maximum value
of $m_s/\hat m$ as a function of the $\pi NN$
coupling constant~\cite{Knecht96}. Some corresponding numbers are summarized in
Table~\ref{tab:sumrule}. If our previously obtained high value of
$g_{\pi NN}$~\cite{Eri95} is confirmed, then a value of $m_s/\hat m$ of 25, as
required by a large value of quark condensate, will be excluded. However,
lower $g_{\pi NN}$ allow this value. These conclusions are dependent on the
precise experimental determination not only of the $\pi NN$ couplings but
also of the $K\Lambda N$ and $K \Sigma N$ couplings. On the theoretical
side, the evaluation of corrections of the order of $m^2_{quark}$, to the
sum rule, should also be performed.

\begin{table}[h]
\caption[h]{Maximum values of  $m_s/\hat m$ as predicted by the
Dashen-Weinstein sum rule, if $|g_{K \Sigma N}| < 7$, as function of
 $g_{\pi NN}$. }

\begin{tabular}{lcccc}
    Source                        &            
    $g_{\pi NN}$                                     &
   $g^2_{\pi NN}/4 \pi $ &   $f^2_{\pi NN}/4 \pi $ &   $(m_s/\hat m)_{max}$ \\
\hline
$np \to pn$: Difference method~\cite{Eri95} &
   $13.55 \pm .14$                    &
$14.62 \pm .30$&   $.0810 \pm .0020$    &            $9.7 \pm 4.3 $    \\
 $\pi^{\pm} p \to \pi^{\pm} p$ ; Dispersion relation~\cite{Koc80}  &
   $13.40 \pm .08$  &
$14.30 \pm .20$  &    $.0790 \pm .0010$ &            $12.3 \pm 5.1$    \\
$\pi^{\pm} N \to \pi^{\pm} N$; GMO sum rule~\cite{Arn94b}  &
   $13.14 \pm .07$       &
$13.75 \pm .15$  &   $.0760 \pm .0008$  &               $22 \pm 10$    \\
$NN \to NN$; PWA~\cite{Sto93a}&
   $13.06 \pm .03$                         & 
$13.58 \pm .05$  &   $.0750 \pm .0003$  &               $28 \pm 12$    \\
\end{tabular}
\label{tab:sumrule}
\end{table}

\section*{NEUTRON PROTON CHARGE EXCHANGE DATA ANALYSIS}

It is very striking that the $np$ unpolarized charge exchange cross sections in
a very large range of energies from about 100 MeV to several GeV, have similar
shape and normalization (in the laboratory system). These data contain
essentially the same physical information as far as the extrapolation to the
pion pole is concerned. Here we shall concentrate our analysis to new precise
data at 162 MeV~\cite{Rahm97} consisting of an extension from
$\theta_{C\!M}=72^\circ$ to $120^{\circ}$ of our previous backward 
measurement~\cite{Eri95}. This allows to improve the absolute normalization
to about $\pm 2\%$. A study of the present $np$ data base~\cite{Blo}, shows
that there are two main families with respect to the angular shape. The first
one is dominated by the Bonner {\it et al.} data~\cite{Bon78}, which have a
flattish angular distribution at backward angles. The second one, which
includes our measurements and the H\"{u}rster {\it et al.}~\cite{Hur80}
data, have a steeper angular shape. The total c.m. cross sections can be
defined in terms of the five amplitudes $a$, $b$, $c$, $d$, $e$~\cite{Bys78}
as

\begin{eqnarray*}
\frac{\rm d\sigma}{\rm d\Omega}(q^2) & = &
      \frac{1}{2}(|a|^2+|b|^2+|c|^2+|d|^2+|e|^2) \\
& = &\frac{1}{2}\left[\frac{1}{2}(|a+c|^2+|a-c|^2) + 
      \frac{1}{2}(|b+d|^2+|b-d|^2)+|e|^2\right];\\
\label{eq:a2b2c2d2e2}
\end{eqnarray*}
where $q^2$ is the squared momentum transfer from the neutron to the 
proton.

\begin{figure}[h]
\begin{center}
\epsfig{figure=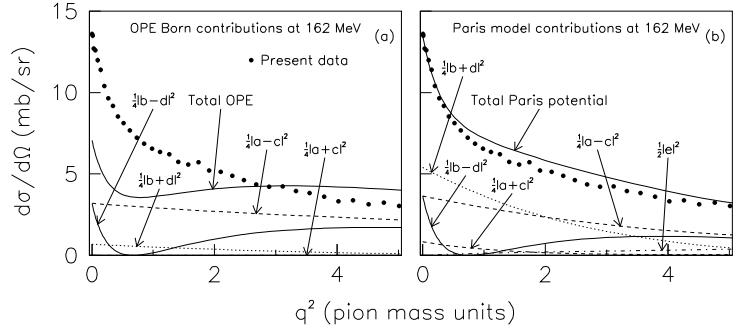,width=10cm}
\end{center}
\caption{Contributions to the $np$
cross section at 162 MeV of combinations of the amplitudes $a$,
$b$, $c$, $d$, $e$ of eq.~\protect\ref{eq:a2b2c2d2e2}. a) for the
regularized pion Born terms b) for the Paris potential model.}
\label{fig:OPE}
\end{figure}

In order to understand the qualitative contributions of pion exchange, we have
chosen the regularized pion Born amplitudes of Ref.~\cite{Gib94} with the
r-space $\delta$-function subtracted. This ensures a non-zero cross section at
$180^{\circ}$. The different components for this Born pion terms, and for the
more realistic Paris potential, are then displayed in 
Figs.~\ref{fig:OPE}a and \ref{fig:OPE}b
, respectively. The combination $|b-d|^2$, which contains the
entire pion pole term, is for the Paris potential remarkably close to that of
the Born term, particularly at small $q^2$. The term $|a+c|^2$ is very small in
both cases and the more important  $|a-c|^2$ terms are again very similar. The
simple structure of the term which contains the pion pole gives considerable
confidence that the extrapolation can be achieved realistically.


\section*{EXTRAPOLATION TO THE PION POLE}

The basic idea to extrapolate to the pion pole is to construct a 
smooth physical function, the Chew function~\cite{Che58},

\begin{equation}
y(x) = \frac{sx^2}{m_{\pi}^4g_R^4} 
       \frac{{\rm d}\sigma}{{\rm d}\Omega}(x) = 
       \sum _{{\rm i=0}}^{{\rm n-1}}a_ix^i.
\label{eq:CHEW}
\end{equation}
Here $s$ is the square of the total energy and $x = q^2+m_{\pi}^2$. At 
the pion pole $x=0$, the Chew function gives  
$y(0) \equiv a_0 \equiv g^4/g_R^4$,
$g$ being the pseudoscalar coupling constant related to the 
 pseudovector coupling by $f = (m_{\pi}/2M_p)g$. The
quantity $g_R^2$ is a reference scale for the coupling chosen for convenience.
The model-independent extrapolation 
requires accurate data with absolute normalization of the differential
cross section. If the differential cross section is incorrectly 
normalized by a factor $N$, the extrapolation determines $\sqrt{N}g^2$. 

The Difference Method, which we introduced to obtain a substantial 
improvement~\cite{Eri95}, is based 
on the Chew function, but it recognizes that a major part in the cross 
section behaviour is described by models with exactly known values for 
the coupling constant. It applies the Chew method to the 
{\it difference} between the function $y(x)$ obtained from a model and 
from the experimental data, i.e., 
\begin{equation}
y_M(x)-y_{exp}(x) = \sum_{{\rm i=0}}^{{\rm n-1}} d_ix^i
\label{eq:Difference}
\end{equation}
with $g_R$ of eq.~\ref{eq:CHEW} replaced by the model value $g_M$. 
At the pole
$y_M(0)-y_{exp}(0) \equiv d_0 \equiv (g_M^4-g^4)/g_M^4$.
This  should diminish systematic extrapolation errors and remove a 
substantial part of the irrelevant information at large  momentum 
transfers.

In our work we have explicitly shown, using `pseudodata' generated from
models in common use	including the Nijmegen potential~\cite{Sto94}, that
we can reproduce the input coupling constants of the models to a precision
less than 1\%. We have grouped the data into a ``reduced range'',
$0 < q^2 <4\ m_{\pi}^2$ with 31 data points and a "full range",
$0 < q^2 < 10.1\ m_{\pi}^2$ with 54 data points. The reduced range is the
range of the data available for the analysis in our previous
work~\cite{Eri95}. This allows to check the sensitivity and stability of the
extrapolation to a particular cut in momentum transfer and to verify that it
is the small $q^2$ region that carries most of the pion pole information.

\begin{figure}[h]
\begin{center}
\epsfig{figure=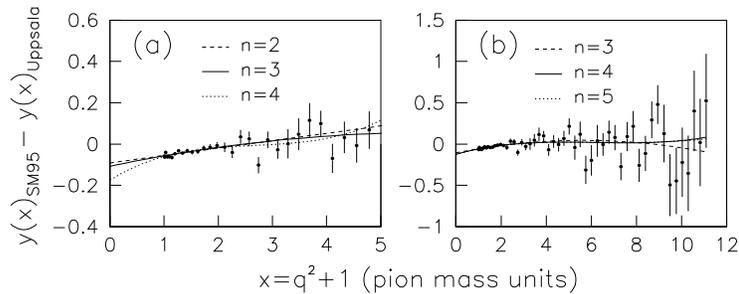,width=10cm}
\end{center}
\caption{Extrapolations of the Chew function $y(q^2)$ to the pion pole
at 162 MeV with the Difference Method using PWA SM95 as comparison
model, different order of polynomials and different intervals in
$q^2$. The left panel uses the reduced range $0 < q^2 < 4\ m_{\pi}^2$;
the right panel uses the full range $0 < q^2 < 10.1\ m_{\pi}^2$.}
\label{fig:diff}
\end{figure}


The Difference Method requires only a few terms in the polynomial expansion in
favorable cases, and this gives a small, statistical extrapolation error. The
similarity between the angular distributions from models and the experimental
data is exploited, particularly for large $q^2$. This incorporates substantial
additional physical information without introducing any model dependence. We
apply the method using three comparison models: the Nijmegen
potential~\cite{Sto94}, the Nijmegen energy-dependent PWA NI93~\cite{Sto93b} as
well as  the Virginia SM95 energy-dependent PWA~\cite{Said,Arn95}. The result,
for this last case, is shown in
Fig.~\ref{fig:diff} 
for the reduced and full
ranges of data. In all cases the extrapolation to the pole can be made easily
and already a visual extrapolation gives a good result. The polynomial fits
cause no problem as long as the data are not overparametrized. The resulting
$g^2=14.52 \pm 0.26$ is consistent with our previous finding~\cite{Eri95}.

Subsequent to our first publication~\cite{Eri95} Arndt {\it et
al.}~\cite{Arn95} subjected a major part of the $np$ charge exchange cross
section data to an analysis using the Difference Method at energies from 0.1
to 1 GeV. They found an average value 13.75 using SM95 as comparison model.
Their individual results show a considerable scatter of approximately
$\pm10\%$. This appears to come from the quality of the data. In particular,
the deduced $g^2$ shows systematic trends with energy as can be seen in
Fig.~3
for the Bonner data leading to an increase of $g^2$ with
energy. Note that for energies above 400 MeV the slope of the data at large
angle is as steep as that of the Uppsala data.


\parbox{8cm}{
\begin{center}
\epsfig{figure=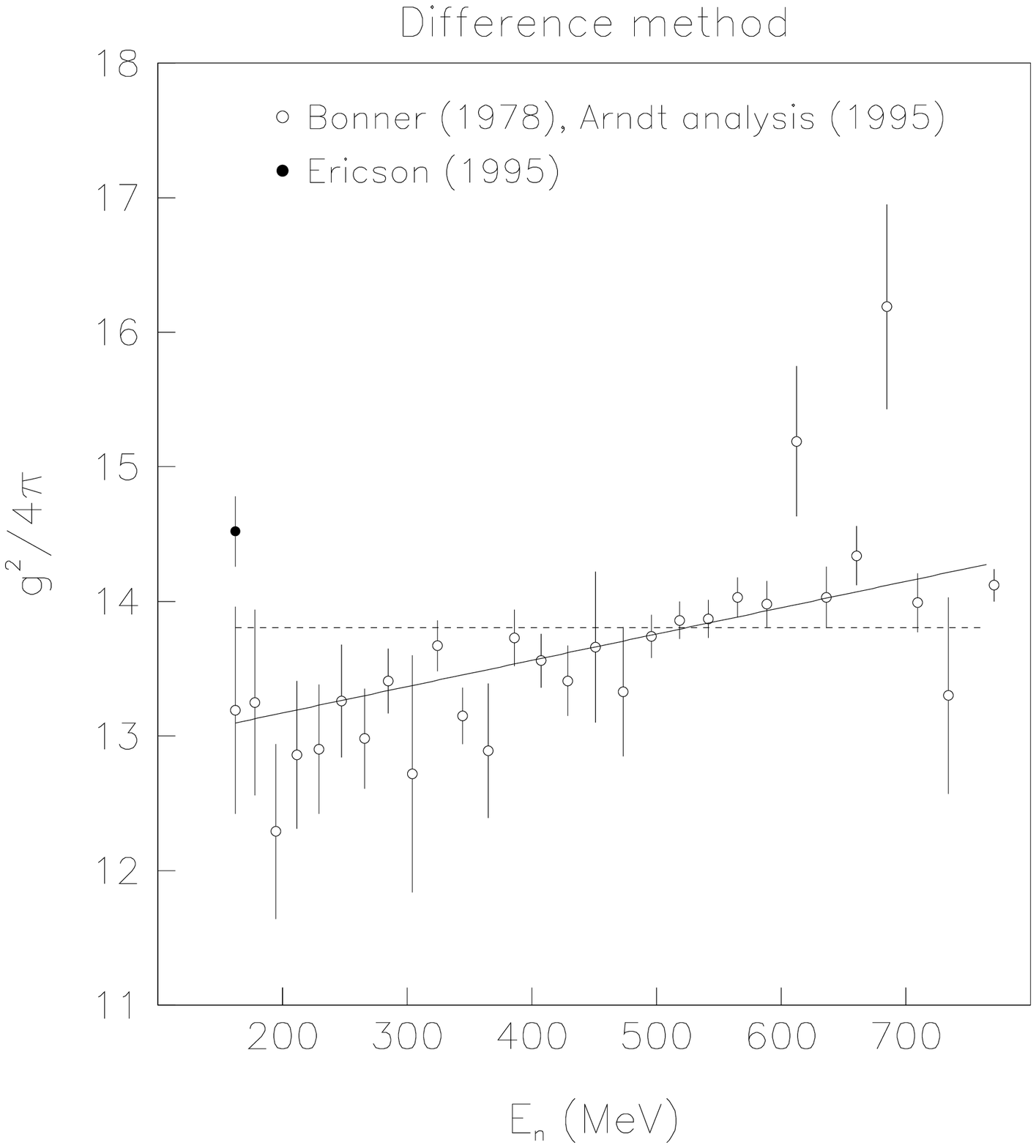,width=8cm,height=8cm}
\end{center}}
\parbox{7cm}{\vspace*{5.0cm}
\noindent
\parbox{5.2cm}
{\small \setlength{\baselineskip}{2.6ex} {\bf Figure.\ 3.}
$g^2$ values obtained~\protect\cite{Arn95} with the Difference method
applied to the Bonner data~\protect\cite{Bon78} using the PWA SM95 as
comparison model. The dotted line is the average $g^2$ when applied to many
data from .1 to 1 GeV, the solid line a linear fit to the Bonner's $g^2$.}}

\section*{CONCLUSIONS}

We have seen that there exists a spread of 7 \% for the value of the $\pi NN$
coupling constant which can have important consequences on our present
understanding of QCD. Here we have  shown that using the most
accurate extrapolation method, the Difference Method, on high precision $np$
differential cross section measurements at 162~MeV in the angular range
$72^\circ-180^\circ$ one can obtain a precise value of this coupling, namely
$ \sqrt{N} g^2 = 14.52 \pm 0.13$ with a systematic error 
of about $\pm 0.15$ and a normalization uncertainty of $\pm 0.17$.
We have no difficulty in reproducing the input coupling constants of models 
using equivalent pseudo-data. The practical usefulness of the method, 
its precision and its relative insensitivity to systematics appear to be 
in hand without serious problems. The data were normalized using the
total $np$ cross section, which is one of the most accurately known cross
sections in nuclear physics, together with a novel approach, in which the
differential cross section measurement was considered as a simultaneous
measurement of a fraction of the total cross section.
It was found that, in the angular region $150^\circ-180^\circ$, our data are
steeper than those of the large data set of Bonner {\it et al.}~\cite{Bon78}
below about 400~MeV. This steeper behaviour, leading to a high value of $g^2$,
should be confirmed and a dedicated $np$ charge exchange precision experiment at
200 MeV with a tagged neutron, to allow an absolute measurement, is going to be
performed at IUCF.

\bibliographystyle{unsrt}

\end{document}